\documentclass[12pt]{article}

\newcommand{\be}{\begin{equation}}
\newcommand{\ee}{\end{equation}}
\newcommand{\ba}{\begin{eqnarray}}
\newcommand{\ea}{\end{eqnarray}}

\begin{document}
\begin{center}
{\huge \bf The Effect of Accumulation \\ of Excess Energy of a Body\\ in Gravitational Compression }\\[2mm]

{ S.~S.~Gershtein, A.~A.~Logunov,  M.~A.~Mestvirishvili \\}
       Institute for High Energy Physics, Protvino, Russia\\[6mm]
\end{center}

\begin{abstract}
It is shown that a gravitational compression of a spherical body
results in an infinite growth of the energy of a body as its radius 
comes close to $GM/c^2$. This gives rise to a negative 
defect of mass and, due to an instability, to an expansion or to an explosion.
A rigorous proof of the above statement can be obtained
within the General Relativity in the harmonic coordinates.
\end{abstract}

The Hilbert-Einstein equation in arbitrary coordinates has the form
[1]
\be
(-g)(T^{ik}+t^{ik}_{g})=\frac{\partial h^{ikl}}{\partial x^l},
\ee
\be {\mbox{where}}\;\;\ h^{ikl}=\frac{\partial}{\partial x^m}\lambda^{iklm},\;\;\;\ 
\lambda^{iklm}=\frac{1}{16\pi}(-g)(g^{ik}g^{lm}-g^{il}g^{km}).
\ee
We use the system of units in which $G=c=1$. From formula (2) it follows that
\be 
h^{ikl}=-h^{ilk}.
\ee
Now, Eqs. (1) and (3) imply the conservation law
\be 
\frac{\partial}{\partial x^k}(-g)(T^{ik}+t^{ik}_{g})=0,
\ee
which is valid in an arbitrary reference frame. Here
$t^{ik}_{g}$ is the Landau-Lifshitz pseudotensor [1]:
\begin{eqnarray}
(-g)t^{ik}_{g}=\frac{1}{16\pi}\{\partial_l\;\tilde g^{ik}\partial_m\;\tilde g^{lm}-
\partial_l\;\tilde g^{il}\;\partial_m\;\tilde g^{km}+\frac{1}{2}\tilde g^{ik}\tilde g_{lm}
\partial_p\tilde g^{ln}\partial_n \tilde g^{pm}-
\nonumber
\\
-(\tilde g^{il}\tilde g_{mn}\partial_p\tilde g^{kn}\partial_l\tilde g^{mp}+
\tilde g^{kl}\tilde g_{mn} \partial_p\tilde g^{in}\partial_l\tilde g^{mp})+
\nonumber
\\
+\tilde g_{lm}\;\tilde g^{np}\partial_n\;\tilde g^{il}\partial_p\;\tilde g^{km}+\frac{1}{8}
(2\tilde g^{il}\;\tilde g^{km}-\tilde g^{ik}\;\tilde g^{lm})
(2\tilde g_{np}\;\tilde g_{qr}-\tilde g_{pq}\;\tilde g_{nr})
\partial_l\tilde g^{nr}\;\partial_m\;\tilde g^{pq}
\},
\end{eqnarray}
$
\mbox{and}\;\; \tilde g^{mn}=\sqrt{-g}g^{mn};\;\;\;\;m,n=0,1,2,3.
$
It should be noted that, in the Relativistic Theory of Gravitation (RTG), 
the conservation law (4) is valid only in an inertial reference frame 
in the Galilean coordinates. 
In what follows, we use the pseudotensor for calculations in General Relativity
(GR) only in the harmonic coordinates.
In view of formula (1), the Landau-Lifshitz pseudotensor
outside of a body coincides with the Fock pseudotensor
and formula (4) implies conservation of the following integrals
of motion [1]:
\be 
P^i=\int(-g)(T^{ik}+t^{ik}_{g})dS_k.
\ee
Performing an integration over the hypersurface $x^0=const$, 
we arrive at the integral over three-dimensional space,
\be 
P^i=\int(-g)(T^{i0}+t^{i0}_{g})dV.
\ee

Our considerations rely heavily on the Hilbert causality
principle, however, it receives only a little attention in the literature.
For this reason, we give an extended quotation from the Hilbert paper
[2]: 

``\textit{So far we considered all coordinate systems 
$x_s$ obtained from some system by an arbitrary coordinate 
transformations as equivalent.
This arbitrariness must be limited if we
consider that two points located on the same time line
remain in casual relationship to each other and, therefore,
they cannot go over into two equal-time points
under a proper coordinate transformation.}

Assuming that $x_0$ is the 
\textbf{proper-time coordinate}, we 
formulate the following definition.
A system of space-time coordinates
is named a \textbf{proper} coordinate system
if, in addition to the condition $g<0$,
the inequalities
$$
g_{11}<0,\;\;\;
\left |
\begin{array}{cc}
g_{11} & g_{12}\\
g_{21} & g_{22}
\end{array}
\right |
> 0, \;\;\;
\left |
\begin{array}{ccc}
g_{11} & g_{12} & g_{13}\\
g_{21} & g_{22} & g_{23}\\
g_{31} & g_{32} & g_{33}
\end{array}
\right |
< 0, \;\;\;
g_{00} > 0.\eqno{(\alpha)}
$$
are valid.
\textit{A transformation which
maps a proper system of space-time coordinates 
into a proper system of space-time coordinates 
is named a \textbf{proper} coordinate transformation''.}
Some later Hilbert notes:
\textit{``Since under a proper
coordinate transformation the time line goes over into time line, 
two world points located at the same time line
never go over into two world points with the same value
of time coordinate; that is, they cannot be associated 
with simultaneous events''...} 
\textit{``In the gravitational field, a single mass point 
moves along geodesic, which represents a time line.
Light moves along a null geodesic.}

\textit{A world line, which is a trajectory of a test particle,
must be a time line. For this reason, we can always 
go over into rest frame of this particle using 
\textbf{proper} coordinate transformations''.}

\textit{The Hilbert conditions  $(\alpha)$ 
on the metric coefficients are indispensable.}

Hilbert emphasized this:
\textit{``We see that the fundamental concept of
cause and effect underlying the causality principle
does not give rise to inconsistencies when we
add the inequalities $(\alpha)$ to our main equations,
that is, if we restrict our consideration to
the \textbf{proper} space-time coordinates}''.

Hilbert causality conditions keep the metric signature
$(+---)$ invariant under proper coordinate transformations.
From the inequalities $(\alpha)$ it is seen that the Hilbert
causality conditions put no limit on the metric 
coefficients $g_{0\alpha}$.

The Hilbert causality principle is of fundamental importance.
This means that only those solutions of the Hilbert-Einstein 
equations make physical sense in General Relativity, 
which satisfy the Hilbert causality conditions.



It was shown [3] that the GR equations with
regard of the Hilbert causality principle
rule out a possibility of the gravitational collapse.
However, behavior of the radius $r_0(t)$ of a body
in gravitational compression has not received sufficient 
attention. In this work, we study its behavior
by considering the integral of motion in harmonic coordinates,
which are privileged coordinates as it was shown by
academician V.A.~Fock.

In the GR, in the case of non-static spherically symmetric body,
an interval outside of the body is static. 
In the harmonic coordinates it has the form [4]
\be 
ds^2=\frac{r-M}{r+M}dt^2-\frac{r+M}{r-M}dr^2-(r+M)^2
(d\theta^2+\sin^2\theta d\phi^2).
\ee

Changing from the spherical to the Cartesian coordinates, we arrive at
$(\alpha,\beta=1,2,3)$ 
\be 
g_{00}=\frac{r-M}{r+M},\;\;\;g_{0\alpha}=0,\;\;\;
g_{\alpha \beta}=-(1+\frac{M}{r})^2\delta_{\alpha \beta}-\frac{r+M}{r-M}\;
\frac{M^2}{r^4}x_{\alpha}x_{\beta}.
\ee
From these expressions we obtain [4]
\be 
\sqrt{-g}g^{\alpha\beta}=\tilde{g}^{\alpha\beta}=-\delta_{\alpha\beta}+M^2\frac{x_\alpha x_\beta}{r^4},\;\;\;
\tilde{g}^{00}=\frac{(r+M)^3}{r^2(r-M)},\;\;\;
\tilde{g}^{0\alpha}=0.
\ee
From the above formulas it follows that the coordinates $x_\alpha$ 
are harmonic [4],
\be 
\frac{\partial \tilde{g}^{\alpha\beta}}{\partial x^\beta}=0.
\ee

Following the ideas of paper [5], we rearrange the integral of motion (7) 
to the form
\be 
P^0=\int_{V_0(t)}(-g)(T^{00}+t^{00}_{g})dV+\int_{V_1(t)}(-g)t^{00}_{g}dV,
\ee
where $V_0(t)$ is the interior of the body 
and $V_1(t)$ is the (infinite) domain outside the body. 
An integration over the domain $V_1(t)$ with the use of harmonic coordinates
$x_\alpha$ gives
\be 
\int_{V_1(t)}(-g)t^{00}_{g}dV=-M^2\biggl(\frac{4}{r_0(t)-M}-\frac{1}{2r_0(t)}+
\frac{M}{r^{2}_{0}(t)}+\frac{M^2}{2r_{0}^{3}(t)}\biggr).  
\ee
The integral of motion (7) is readily calculated
by reducing to an integral over a surface [1]
\be 
P^0=M.  
\ee

Substituting expressions (13) and (14) in formula (12), we obtain 
\be 
\int_{V_0(t)}(-g)(T^{00}+t^{00}_{g})dV=M+M^2\biggl[\frac{4}{r_{0}(t)-M}-\frac{1}{2r_{0}(t)}+
\frac{M}{r_{0}^{2}(t)}+\frac{M^2}{2r_{0}^{3}(t)}\biggr].
\ee
This formula was derived earlier in the Relativistic Theory of
Gravitation (RTG) due to the existence of the energy-momentum
\textit{tensor} of the gravitational field [5].
The physical solution is obtained by "sewing"
of the non-static spherically symmetric solution inside of a body
with the solution (8) in the exterior domain.
Since it satisfies the Hilbert causality conditions,
the volume of a body in gravitational compression is always finite
and, therefore, the energy of a body is also finite [6].

Finiteness of the energy of a body in gravitational compression,
as it follows from formula (15),
rules out a possibility of gravitational collapse.
Thus we conclude that the quantity  $GM/c^2$ is of fundamental importance
--- it gives the radius that cannot be approached by a body of mass $M$:
\be
r_0(t)>M.
\ee
Therefore, a massive body cannot be pointlike.
A calculation of the value of $t^{ik}_{g}g_{ik}=t_g$
in the GR in harmonic coordinates gives [5]
\be
32\pi t_g=-\frac{r_{g}^{2}(24r^2-16rr_g+5r_{g}^{2})}{4r^2(r-M)(r+M)^3};\;\;\;\;
r_g=2M.
\ee

The value of $t_g$ must be limited in order to sew solutions in
the interior and exterior domains, i.e. it is necessary
\be
r>M,
\ee
in agreement with the condition (16).
In RTG, the quantity $t_g$ is an invariant of the gravitational field [5].

\textbf{From formula (15) it follows that,
when thermal and thermonuclear sources of energy of a star are exhausted, 
a star of large mass $(M>3M\odot)$ accumulates an excess of positive energy
from gravitational compression, which results in a negative mass defect.
This, in its turn, stops compression and gives rise to a subsequent
radial expansion of a body. Due to an instability, such 
expansion can be of explosive type providing the validity
of inequality (16). After multiple repetition of such process 
it may be that only stable objects remain.}

Our conclusions agree with the empiric model proposed
by Ambarzumian in 1960. In this model (see [7]), ``superdense matter
evolves into a diffuse state; positive-energy states can
explode and transform from dense to diffuse state''.
For more detail, special computations are needed.
States with excess energy as well as other
positive-energy solutions were considered in [7],[8].
 
Provided that the Landau--Lifshitz pseudotensor
is calculated in the GR in harmonic coordinates,
one obtains neither gravitational collapse nor black holes.



We are grateful to V.A.~Petrov, A.P.~Samokhin and N.E.~Tyurin for useful discussions.

\end{document}